\documentclass[epj]{svjour}

\usepackage[super,compress]{cite}
\usepackage{graphicx}

\usepackage{braket}
\usepackage{multirow}
\usepackage{dsfont}
\usepackage{mathtools}
\usepackage{xcolor}
\usepackage[normalem]{ulem}

\renewcommand{\vec}[1]{\mbox{\boldmath$#1$\unboldmath}}

\begin{document}

  \title{Large $N_c$ QCD phase diagram at $\mu_B=0$.}
  
 \author{T.~D.~Cohen\inst{1} \and  L.~Ya.~Glozman\inst{2}}
%
\institute{Department of Physics and Maryland Center for Fundamental Physics, University of Maryland,\\ College Park, MD 20742 USA\and
Institute of Physics, University of Graz, A-8010 Graz, Austria}
%
%
 
\abstract{
Lattice studies suggest that at zero baryon chemical potential and increasing temperature there are three characteristic regimes in QCD that are connected by smooth analytical crossovers: a hadron gas regime at $T < T_{ch}\sim 155$ MeV, an intermediate regime, called stringy fluid,  at $T_{ch} < T < \sim 3 T_{ch}$,  and a   quark-gluon plasma regime at higher temperatures.   These regimes have been interpreted to reflect different approximate symmetries and effective degrees of freedom. In the  hadron gas the effective degrees of freedom are hadrons and  the approximate chiral symmetry of QCD is spontaneously broken. The intermediate regime has been interpreted as lacking
spontaneous chiral symmetry breaking along with the emergence of new approximate symmetry,  chiral spin symmetry, that is not a symmetry of the Dirac Lagrangian, but is a symmetry of the confining part of the QCD Lagrangian.  While the high temperature regime is the usual quark-gluon plasma which is often considered to reflect ``deconfinement'' in some way.
This paper explores the behavior of these regimes of QCD as the number of colors in the theory, $N_c$, gets large.  In the large $N_c$ limit  the theory
is  center-symmetric and notions of confinement and deconfinement are unambiguous.
 The energy density is ${\cal O}(N_c^0)$ in the meson gas, ${\cal O}(N_c^1)$ in the intermediate regime and  ${\cal O}(N_c^2)$ in the quark-gluon plasma regime.
 In the large $N_c$ limit these regimes may become distinct phases separated
 by first order phase transitions. The intermediate phase has the peculiar feature that  glueballs should exist and have properties that are unchanged from what is  seen in the vacuum (up to $1/N_c $ corrections), while  the ordinary dilute gas of mesons with broken chiral symmetry  disappears  and approximate chiral spin symmetry should emerge.  
}

\maketitle

\section{Introduction}

In the early days of QCD there was a widespread belief that QCD had a deconfinement phase transition at the Hagedorn temperature in which the system went from a dilute hadron resonance gas to a weakly interacting quark-gluon plasma\cite{Hag,Cab,Col}.  However, it was recognized long ago that this picture is too simple. Moreover, RHIC experiments have established that the matter created within a fireball while very different from a dilute hadron gas, is also very different from a weakly coupled quark-gluon plasma (QGP)\cite{BNL1,BNL2,BNL3,BNL4,BNL5}. These properties have been reconfirmed at higher temperatures at the LHC. A prominent characteristic of this matter is a small value $\eta/s \sim 0.2$ (where $\eta$ is the shear viscosity and $s$ the entropy density) both at RHIC and LHC temperatures \cite{H1,H2};  this implies that this matter behaves as a strongly coupled highly collective liquid with a very small mean free path of the effective constituents. Another indication of the strongly-coupled nature of the matter within the fireball is a modification of a jet that propagates through the fireball\cite{jet}. It was often assumed that this matter  is a system of deconfined strongly interacting quark and gluon quasiparticles which was reflected in the name sQGP.

In parallel with the experimental activity,  hot QCD matter in the equilibrium at a given temperature was investigated on the lattice. At zero chemical potential a smooth analytic crossover around the chiral restoration pseudocritical temperature $T_{ch} \sim 155$ MeV \cite{A1,A2} was established by a few groups\cite{ch1,ch2}  and  it was widely believed that  a deconfinement crossover occurs in the same temperature region. However,  it has become increasing clear that this belief was erroneous \cite{G1}.  

In retrospect, it  is hardly surprising that the old belief that deconfinement and chiral restoration happened together was erroneous: general theoretical principles suggest that there is no rigorous way that the two can be directly connected.  Note that both spontaneous chiral symmetry breaking and deconfinement are associated with symmetries that are exact only in particular limits of the the theory and that the limits associated with the two are incompatible.  Chiral symmetry  becomes exact in the limit of massless quarks.  In contrast, the only strict definition of deconfinement at our disposal is  QCD with infinitely heavy quarks (where QCD degenerates into Yang-Mills theory); in that case the action has an $Z_{N_c}$ center symmetry.  This $Z_{N_c}$ center symmetry becomes broken in the high temperature phase; its breaking is associated with the disappearance of a linearly-rising confining effective potential between fundamental color charges and can be taken as a clean definition for deconfinement.  The difficulty is that in the limit where chiral symmetry is cleanly defined, there is no clean way to define deconfinement and conversely.  This strongly suggests that at  theoretical level there is no unambiguous way to relate the two phenomena.  Of course in real QCD the quarks are neither massless nor infinitely heavy and both the notion of spontaneous chiral symmetry breaking and deconfinement are somewhat fuzzy notions.  

In this respect, the large $N_c$ limit of QCD with massless quarks may help clarify issues theoretically since the theory is manifestly chirally symmetric and at the same time, explicit violations of $Z_{N_c}$ center symmetry become increasingly suppressed as the $N_c \rightarrow \infty$ is approached; in the limit  $Z_{N_c}$ center symmetry becomes exact and morphs into a continuous $U(1)$ symmetry rather than a discrete symmetry.  This paper will explore this large-$N_c$ theory for insights into what is going on.  

Before exploring the large $N_c$ world, it is important to look in a bit more detail at the behavior of real-world  QCD with $N_c=3$ and small, but not zero quark mass.  It has become increasingly clear\cite{G1} that the crossovers associated with chiral restoration (as indicated, for example, by a rapidly dropping chiral order parameter, leading to a maximum in the chiral susceptibility at $T_{ch}$) and the cross-over into a QGP regime  occur at essentially different temperatures; the latter above $\sim 3T_{ch}$.    Moreover there appears to be very interesting behavior in the intermediate regime. 

Tools to understand this behavior were developed starting almost ten years ago.  The initial hint involved an artificial truncation  of the near-zero modes of the Dirac
operator in lattice calculations \cite{D1,D2,D3,D4}; these studies contained evidence of an approximate emergent  symmetry \cite{G2,G3} in this artificial regime.  It is emergent in the sense that while there is  evidence for the symmetry in this regime, the symmetry is larger than the symmetry of the QCD Lagranigan.    This emergent symmetry is the $SU(2)_{CS}$ ``chiral spin''
symmetry; it has as a subgroup $U(1)_A$.   Its flavor extension to $N_f$ light flavors is $SU(2N_F)$, which includes $SU(N_F)_L \times SU(N_F)_R \times U(1)_A$. These are symmetries of the
color charge and of the electric part of the QCD Lagrangian \cite{G1}.

Since the confining dynamics of non-abeliean gauge theories has often been associated with the formation of color-electric flux tubes, it is not unnatural to identify the color-electric interaction of QCD as the ``confining interaction''.   In the remainder of this paper, we will refer to this part of  the QCD Lagrangian or Hamiltonian in this way. We recognize that it is possible to object to this identification as treating the complex ideas associated with confinement in a very simple way.  Nevertheless it is helpful to make this identification, minimally  for ease of communication.  Moreover, it is useful conceptually to take seriously the identification and see where it leads.  This identification of the confining interaction is interesting since the confining interaction, so identified, is invariant under chiral spin---while the color-magnetic interaction as well as the quark kinetic terms are not.  

A clear implication
of these truncation studies is that the confining interaction in QCD is distributed among all modes
of the Dirac operator, not only in the near zero modes that are responsible for  $SU(2)_L \times SU(2)_R \times U(1)_A$
breaking\cite{G1,G2,G3}. To the extent that the identification of the confining interaction is valid, this result strongly suggests that confinement and spontaneous chiral symmetry breaking are distinct phenomena that are not directly related, which, as noted above, was typically assumed in the early days of QCD.  

To the extent that truncation studies capture a general feature of QCD,  there are obvious implications for hot QCD. Recall that as the system is heated,  the density of near-zero Dirac modes is suppressed.  The Banks-Casher relation\cite{bc} relates the chiral condensate  to this density as a direct proportionality so that  above the chiral
symmetry restoration crossover where this density becomes suppressed, one might naturally expect the system to behave much like the system in which the near-zero Dirac modes were removed by hand.   Moreover, one might expect that along with chiral restoration, an approximate  chiral spin
symmetry and its extensions  could naturally emerge since it was seen in the artificial regime \cite{G4}. This scenario was verified to occur in  lattice simulation  in $N_F=2$ QCD with the
domain wall Dirac operator\cite{R1,R2,R3}, where
distinct multiplets of the above groups were observed in spatial and temporal correlators above $T_{ch}$
but below roughly $3T_{ch}$. Similar conclusions have been obtained very recently in $N_F = 2+1+1$ QCD \cite{Chiu}.  Since the confining part of the QCD interaction respects these symmetries there is no reason to believe that confinement dynamics need also be suppressed in this regime.  Thus one would expect that although (approximate) chiral symmetry has been restored to good approximation, the system would still be  in a confining regime.

From these lattice studies, three regimes of QCD were identified with clearly distinguishable approximate symmetries:
below $T_{ch}$ spontaneously broken chiral symmetry ,  chiral restoration with an approximate emergent $SU(2)_{CS}$ (as well as $SU(2 N_f)$ symmetries)  at $T_{ch} < T < \sim 3 T_{ch}$ and with just the usual   chiral symmetries (without spontaneous breaking)  at higher temperatures.  A simple way to interpret this pattern, is to suggest that after the initial chiral crossover, the dominant interaction is the confining one, which respects these enlarged symmetries.  These enlarged symmetries are only approximately realized in this regime since the confining interaction is dominant but other interactions 
and quark kinetic terms are present; but to the extent the confining interaction dominates the patterns associated with the enlarged symmetries are clearly discernible.   As the temperature increases towards a deconfined regime the effect of confining interaction weakens relative to the remainder and the enlarged symmetries cease to be observable.

One possible way to interpret these observations in terms of a physical picture is to suggest that at $T_{ch} < T < \sim 3 T_{ch}$, the effective degrees of freedom act like chirally symmetric quarks that  are bound by a  electric field into color-singlet hadron-like objects. Consequently this regime was dubbed
a stringy fluid. Note that these hadron-like clusters are densely packed so that the quark
interchanges between them required by Pauli principle are significant. Still it is a system
with confinement.
At higher temperatures the confining electric field gets screened and one observes a smooth transition to the QGP. Note that the physical picture of a ``stringy  fluid'' is simply a heuristic way  to view things.  The overall pattern of approximate multiplets for correlation functions that are consistent with the symmetries outlined is a physical phenomenon that exists regardless of whether this heuristic picture ultimately turns our to be accurate.

There is additional evidence that the regime $T_{ch} < T < \sim 3 T_{ch}$, is not  described by a simple description in which (quasi)partons are the effective degrees of freedom:  screening masses and equation of state \cite{GPP}. Direct evidence for effective hadron degrees of
freedom above $T_{ch}$ can be seen from  the pion spectral function extracted from the spatial correlators \cite{LP,LP2}.  These indicate 
that while $\pi$ and $\pi'$ excitations become broader with temperature, clear and discernible hadron-like structures  persist in the regime; these structures cease to be discernible above around $3 T_{ch}$.
Other direct evidence of hadronic degrees of freedom above  $T_{ch}$ is the bottomonuim spectrum with
the $1S,2S,3S,1P,2P$ states with similar masses like in  vacuum remain in this regime; they simply become broader with temperature \cite{Lar}. For
review of symmetries and lattice evidences of hadronic origin of the stringy fluid regime see Ref. \cite{G1}.

There is also experimental evidence  for hadron degrees of freedom above $T_{ch}$ obtained at CERN SPS: 
dimuon spectra
coming from the fireball demonstrate a clearly discernible $\rho$ structure; it is broader than in vacuum \cite{NA60}.
An extraction of the fireball temperature at SPS via the black body radiation yields $T \sim 205$ MeV, which is clearly above $T_{ch}$ \cite{SPS2}. 

All this evidence suggests three different regimes in QCD that are connected by smooth crossovers.

This paper will use analysis based on the large $N_c$ limit of QCD and $N_c$ counting rules to attempt to gain insight into the phenomena by asking how one would expect the phenomena to evolve with $N_c$ as $N_c$ gets large. There is a realistic prospect that the physics  becomes significantly more transparent at large $N_c$. 


There are two related reasons that large $N_c$ could provide insight. The first was noted above: as $N_c \rightarrow \infty$, the effects of explicit $Z_{N_c}$ center symmetry breaking in QCD diminishes and becomes an exact $U(1)$ symmetry.  Thus for QCD with massless quarks and in the large $N_c$ limit, it becomes possible in principle to simultaneously and unambiguously identify whether the system is in a regime that is chirally restored and whether it is deconfined; this makes it possible to ask well-posed questions about the nature of these regimes without recourse to heuristic and imprecise 
arguments.\footnote{In the large $N_c$ limit there are two independent clear cut
order parameters, one associated with confinement (the Polyakov loop)
and one for chiral symmetry (the quark condensate). We also
note that the 't Hooft anomaly matching conditions \cite{at} and
the Coleman-Witten theorem for chiral symmetry breaking \cite{cw} do not connect the notion
of confinement to center symmetry.  Importantly the Coleman-Witten theorem  exploits Lorentz invariance and so by construction is restricted only to zero temperature. Similarly, the 't Hooft anomaly matching is valid at $T=0$
and its generalization to $T > 0$ is unclear.}   The second is that thermodynamic observables such as the energy density, $\epsilon$, pressure, $P$ or entropy density, $s$  naturally scale differently with $N_c$ in the various regimes:
In a hadronic gas,
\begin{subequations}
\begin{equation}
    \epsilon_{\rm had} \sim N_c^0 \; \; , \; \; P_{\rm had} \sim N_c^0 \; , \; s_{\rm had} \sim N_c^0 \; , \label{Eq:had scale}
\end{equation}
 in a quark-gluon-plasma regime,
\begin{equation}
    \epsilon_{\rm QGP} \sim N_c^2 \; \; , \; \;    P_{\rm QGP} \sim N_c^2\; , \; s_{\rm QGP} \sim N_c^2 \label{Eq:QGPscale0}
\end{equation}
and assuming that an intermediate regime that is confined (in the sense of having a non-zero string tension and unbroken center symmetry) but without spontaneous  chiral symmetry breaking, survives in the large $N_c$ limit 
\begin{equation}
    \epsilon_{\rm int} \sim N_c^1 \; \; , \; \; P_{\rm int} \sim N_c^1 \;  ,  \; s_{\rm int} \sim N_c^1 \;  . \label{Eq:IntScal}
\end{equation}
\end{subequations}
As the $N_c \rightarrow \infty$ limit is approached these regimes become infinitely separated and thus in the limit must correspond to distinct phases separated by phase transitions.  This in turn makes the regime in question unambiguous.  

Before proceeding, it is worth noting that {\it a priori} there are four main scenarios
for the QCD phase diagram at $\mu_B=\mu_I=0$ (where $\mu_B$ and $\mu_I$ are the baryon and isospin chemical potentials respectively)  for the large $N_c$ limit with massless quarks that appear to be consistent  with what is known about the large $N_c$ world based on theoretical scaling arguments\cite{50years,Witten,tHooft} and lattice studies for systems with large but finite $N_c$.

\begin{enumerate}
\item There are two distinct phases, the hadron resonance gas and the QGP, separated by a single first-order phase transition at large $N_c$; the latent heat is of order $N_c^2$ 
\cite{mlp}. The existence of three qualitatively different regimes observed at $N_c=3$ does not persist as $N_c$ gets large, with the intermediate regime vanishing in the large $N_c$ limit.  \label{s1}
\item There are two distinct phases separated by a single first-order phase transition at large $N_c$; the latent heat is of order $N_c^2$.  The low temperature phase is a conventional hadron phase, while the high temperature phase restores chiral symmetry and spontaneously breaks center symmetry at all temperatures in the regime.   However the high temperature phase does not act like weakly interacting plasma until well above the phase transition and hadron-like structures (as would be  seen in correlation functions) would be present at temperatures near the transition.  There could be a smooth cross-over to a regime in which these hadron-like" structures cease to be discernible as temperature increase. \label{s2}
\item  There are three distinct phases at large $N_c$ with the low-temperature hadronic phase separated from the intermediate phase by a second-order phase transition;  the intermediate  phase separated from the high-temperature QGP phase  by a first order transition with a latent heat of order $N_c^2$.  \label{s3}
\item  There are three distinct phases at large $N_c$ with the low-temperature hadronic phase separated from the intermediate  phase by a first-order phase transition, with a latent heat of order $N_c$; the intermediate phase is separated from the high-temperature  QGP phase by a first order transition with a latent heat of order $N_c^2$.\label{s4}
\end{enumerate}

As will be discussed below the plausibility of the various depends in part on the assumptions one makes about the how QCD dynamics manifests itself in the various regimes.


\section{Large $N_c$ counting rules}  

Before discussing the various scenarios it is useful to briefly remind the reader of standard results of large $N_c$ analysis with an emphasis on the implications for thermodynamics.

Most studies of large $N_c$ QCD are based on studies of correlation functions of local color singlet operators that (too leading order in $1/N_c$) cannot be decomposed into more than one color-singlet\cite{Witten,tHooft,50years}.   The $N_c$ of these correlators can be deduced from scaling of Feynman diagrams that contribute to them.  Two key rules can be deduced that control this scaling.  The first is that connected diagrams composed solely of gluon lines (and sources that connect to gluon lines) scale as $N_c^2$ if the diagram is planar and slower than $N_c^2$ if they are non-planar.  The second is that planar diagrams that contain quarks loops that bound the diagram are suppressed by a factor of $1/N_c$ per quark loop relative to diagrams composed entirely of gluons with addition suppression if the diagram is non-planar or the quark loop fails to bound the diagram.

From the $N_c$ counting of the correlators along with the additional assumption of confinement (in the sense that all physical states of the theory are color singlets)  it is straightforward to deduce the following rules for the $N_c$ scaling of properties\cite{Witten,tHooft,50years} of mesons and glueballs:

  \begin{itemize}
 \item At large $N_c$, properties of mesons and glueballs can be described via an effective tree-level theory;  at leading order vertices scale with $N_c$ according to 
$N_c^{2-n_{\rm g}-\frac12 n_{\rm m}}$, for a vertex with  $n_{\rm m}$ meson lines  and $n_{\rm g}$  gluon lines at the vertex which means that at leading order
\item  Mesons and glueballs masses  scale as $N_c^0$.
 \item Meson decay amplitudes scale asymptotically as $N_c^{-1/2}$ and their widths as $N_c^{-1}$. Glueball  decay amplitudes scale as $N_c^{-1}$ and their widths as $N_c^{-2}$.
  \item  Mesons and glueballs are stable at large $N_c$.
\item These hadrons interact weakly.
   \item   Scattering  amplitudes for meson-meson interactions, glueball-glueball interactions and meson-glueball scale as $1/N_c$, $1/N_c^2$, and $1/N_c^2$ respectively.
\item For every allowed quantum number an infinite number of distinct mesons and  glueballs exist. \cite{Witten}.
\item In terms of $N_c$ counting,  hybrid mesons  with exotic quantum numbers scale at large $N_c$ the same way as  ordinary mesons \cite{Cohen:1998jb}.  
\item At large $N_c$ The OZI rule \cite{Okubo:1963fa,Zweig:1964jf,Iizuka:1966fk} is exact .  
  \item The effects of the axial $U(1)$ anomaly are suppressed. 
      \item  Glueball-meson mixing is suppressed.
  \end{itemize}

As noted above meson and glueball masses are independent of $N_c$ at large $N_c$.  In contrast, as discussed by Witten \cite{Witten,50years} baryon masses scale as $N_c^1$ for the simple reason that a baryon contains at least $N_c$ quarks.  This means that at temperatures of $N_c^0$--the regime of interest in this paper, baryons are suppressed by factors that are exponential in $N_c$ and hence negligible.

As the role of chiral symmetry plays an essential role in the analysis it is useful to note the $N_c$ scaling of several quantities that are central to chiral physics, the pion decay constant---$f_\pi$, the chiral condensate---$\langle \overline{q}q \rangle$, the $\sigma$ term for the $j^{\rm th}$ species of meson---$\sigma_{m_j}$ and the  $\sigma$ term for the $k^{\rm th}$ species of glueball---$\sigma_{m_g}$
\begin{equation}
    f_\pi \sim N_c^{1/2} \; \; , \; \; 
\langle \overline{q}q \rangle \sim N_c^1 \; , \; \; \sigma_{m_j} \sim N_c^0 \; \;  \sigma_{m_g} \sim N_c^0 \; ;
\end{equation}
 the $\sigma$ term of hadron $h$ is given by $m_q \partial m_h/\partial m_q$ where $m_q$ is the average of the light quark masses and $m_h$ is the mass of the hadron and corresponds to the shift in the  matrix element of $m_q \langle \overline{q}q \rangle$ in the hadron state relative to the  vacuum.  Note that the $\sigma$ term cannot be defined unambiguously for a resonance rather than a stable particle, but at large $N_c$ mesons and glueballs are stable.

Related to chiral symmetry at large $N_c$ the $\eta'$ is an ordinary pseudo-Goldstone boson since the effects of the axial $U(1)$ anomaly are suppressed.  

 One aspect of large $N_c$ analysis that plays a key role in what follows is the fact that properties of glueballs and mesons scale in qualitatively different ways. A critical difference is that while meson-meson scattering amplitudes scale as $N_c^{-1}$, meson-glueball (and glueball-glueball) scattering amplitudes scale as $N_c^{-2}$.  Thus although both mesons and glueballs are hadrons with masses of order $N_c^0$ that couple weakly to other hadrons, glueballs couplings are qualitatively weaker.  This can have a profound effect on propagation in a medium with a moderate density of hadrons: the medium will affect meson propagation substantially more than glueballs and this leads to the possibility that an intermediate phase could exist with glueball degrees of freedom unaffected by the phase (neglecting $1/N_c$ corrections) while meson degrees of freedom could be radically altered.  Ultimately the cause of the different scalings can be traced to the fact that local gauge invariant operators with meson quantum numbers must couple to quarks and the number of distinct colors for quarks is $N_c$ while operators that couple to glueballs couple to gluons of which there are $N_c^2-1 \sim N_c^2$

This may seem somewhat counter intuitive if one regards both glueballs and mesons as generic hadrons; in that case one might think that they ought to behave similarly.  Of course, in practice the community has had little if any experience with the behavior of glueballs in the physical world of $N_c=3$, since there are few or no states in nature that are clearly glueballs. (The most plausible reason for this is simply that possible glueballs tend to be heavy.)  Thus any intuition one might have that glueballs are ordinary hadrons similar to mesons is not based on significant experience in the $N_c=3$ world.



\section{The phases at large $N_c$}
\subsection{The hadron gas phase}

The large $N_c$ limit greatly simplifies the nature of the hadronic resonance gas phase.   Since the phase exists only at temperatures of order $N_c^0$, baryons can be neglected.  Moreover at large $N_c$ hadrons (mesons and glueballs) do not interact so the phase should be accurately modeled at any given temperature consistent with the phase by a gas of multiple species of non-interacting hadrons with their various masses in thermal equilibrium\cite{Cohen:2006qd}.  Each species of density and momentum distribution are given by the appropriate Bose-Einstein distribution in the same manner used phenomenologically in hadron resonance gas models\cite{Venugopalan:1992hy,Dashen:1969ep}.  This allows a straightforward calculation of thermodynamic properties such as energy density, entropy density and pressure as a function of temperature as well as the number density of each species of hadron.

\begin{subequations}
 The number density of species $k$,  $n_k(T)$, is given by
\begin{equation}
  n_k(T)  = (2S_k+1)(2I_k+1) \int \frac{{\rm d}^3 p}{(2 \pi)^3} \, \frac{1}{e^{\sqrt{p^2 + m_k^2}/T}-1}  \; ,
\end{equation}
where $s$ and $I$ are the spin and of the species. 
The energy density and pressure of the system are given by 
\begin{equation}
    \epsilon_{\rm had}(T)  = \sum_k (2S_k+1)(2I_k+1) \int \frac{{\rm d}^3 p}{(2 \pi)^3} \, \frac{
\sqrt{p^2 + m_k^2}}{e^{\sqrt{p^2 + m_k^2}/T}-1} 
\label{ed}
\end{equation}
\begin{equation}
P_{\rm had(T)} =  \sum_k (2S_k+1)(2I_k+1) \left ( \frac{m_k^2 T^2}{2 \pi^2} \right ) \,
\sum_{j=1}^\infty \, K_2 \left(\frac{j m_k}{T} \right) 
\end{equation}
where the sum over $k$ represents species of hadron (glueballs and mesons) and $K_2$ is a modified Bessel function.  The entropy density is given by 
\begin{equation}
s_{\rm had(T)}= \frac{\epsilon+P}{T}
\end{equation}
\end{subequations}
It is easy to see that the $\epsilon_{\rm had}(T)$, $P_{\rm had}(T)$ and $S_{\rm had}(T)$ scale as in Eq.~\ref{Eq:had scale} and that he number density of each species of hadrons also scales as $N_c^0$.

In the hadronic  regime,the fractional change in the chiral condensate at temperature $T$ relative to the vacuum is given by \cite{Cohen:2004cd}
\begin{equation}
    \frac{\langle \overline q q \rangle_T}{\langle \overline q q \rangle_{\rm vac} }= \frac{m_\pi^2 f_\pi^2- \sum_{j } n_j(T) \sigma_j }{m_\pi^2 f_\pi^2 }  = 1 - {\cal O}(N_c^{-1}) \label{Eq:chi_shift}
\end{equation}
where the sum is over hadron (meson and glueball) species, $n_j$ is the density of the species and $\sigma_j$ is the sigma term for that species.  Note there is an implicit $m_q \rightarrow 0$ limit on the right hand side of the expression;  it is well defined in the chiral limit  since $m_\pi^2$ and the sigma term are each proportional to the quark mass. The key point is that   since  $n_j$, $\sigma_j$ and $m_\pi^2$ are each of order $N_c^0$ while  $f_\pi^2$ is of order $N_c$, it is apparent that at large $N_c$ , $\frac{\langle \overline q q \rangle_T}{\langle \overline q q \rangle_{\rm vac} }= 1$ and to leading order, the chiral condensate is unchanged from its vacuum value.  


\subsection {The intermediate phase}
From the perspective of large $N_c$ analysis and  given current knowledge there is no guarantee that an intermediate phase that is confined ( in the sense of having a non-zero string tension and unbroken center symmetry) but without spontaneous chiral symmetry breaking exists.  Indeed of the  four scenarios considered in this paper, only two have such a phase.   However, it is important to focus on such a phase as it is extremely interesting.  In particular, if such a phase does exist one can deduce  key properties that the phase  must have at large $N_c$.  The most startling of these is that in this phase glueballs exist and they must interact weakly with themselves and with the quark degrees of freedom.

Recall that all diagrams containing quark loops are suppressed by factors of $1/N_c$.  Thus the difference between every correlation function of local, gauge-invariant  sources containing only gluon fields in QCD  and  the same correlation function in Yang-Mills is of order $1/N_c$ and vanishes in the limit: at large $N_c$ these correlation functions are the same in QCD and Yang-Mills.    However, Yang-Mills in the regime where center-symmetry is unbroken is in a confining phase where all correlation functions are unchanged from their vacuum values (up to  corrections of relative order $1/N_c$).  These correlators  describe a gas of  glueballs that neither interact among themselves nor with any other degrees of freedom in the system.  Thus, without further analysis one knows that if such an intermediate phase exists in the large $N_c$ limit it contains a gas of glueballs that neither interact with each other or with degrees of freedom containing quarks.

On the other hand, by hypothesis the intermediate phase is chirally restored which implies that n-point functions of local gauge-invariant color-singlet sources that include quark-bilenears are radically different from those in the low-temperature hadronic gas phase.   Consider, for example the chiral condensate, the one-point functions of $\overline{q} q$.  Its matrix element is generically of order $N_c^1$ and its vacuum value is indeed of that order.  In the ordinary hadron gas phase,  Eq.~(\ref{Eq:chi_shift}) implies that the change in the chiral condensate relative to the vacuum at $T=0$ is known to be suppressed by a factor of $1/N_c$.  We know that this cannot be the case in the intermediate phase where the spontaneously symmetry breaking does not occur: the chiral condensate has shifted from its vacuum value by an amount of order $N_c^1$ ({\it i.e.} from order $N_c^1$ to zero).  This implies that in the putative intermediate phase, the condition restricting shifts in connected n-point functions of local gauge-invariant single-color trace sources to being of relative order $1/N_c$ compared to the vacuum does not apply: in the intermediate phase shifts of relative order unity in a $1/N_c$ expansion are permitted.

However, the quark kinetic energy term,  a 1-point function of the quark bilinear $\overline{q} \vec{D} \!\!\!\! \slash  q$, is a contribution to the energy density.  If it 
is shifted of relative order unity relative to the vacuum (absolute order $N_c^1)$, then there will be shifts in the energy of order $N_c^1$ as seen in Eq.~(\ref{Eq:IntScal}).

There is a potential caveat to that conclusion.  It remains logically possible (although quite implausible) that the chiral condensate shifts from its vacuum by amounts of absolute order $N_c^1$ but other n-point functions of connected n-point functions of local gauge-invariant single-color trace sources do not.  However, there is a strong evidence that this is not the case and that shifts of order $N_c^1$ are generic.  Consider in particular,
correlators of the quark bilinear sources seen in  chiral
spin symmetry.  The connected two-point functions of these operators are all of order $N_c^1$ in the vacuum and are generically different from one another in the vacuum and the hadron gas phase (excepting correlators connected by isopin).  However in a phase with approximate chiral spin symmetry which, by hypothesis, the intermediate phase has many of these correlators become (nearly) identical---indicating a shift of absolute order $N_c^1$ relative to the vacuum and hadron gas phases.  This strongly suggests that shifts  of connected n-point functions of local gauge-invariant single-color trace sources are generically  of order $N_c^1$.

 Moreover, the pressure is related to energy density via the speed of sound: $\left .\frac{P}{\epsilon} \right |_s= 1/v_s^2$.  Since the speed of sound is bounded from above by unity, one sees that the pressure in such a phase must scale as $N_c$ as well.

 There remains an open question regarding chiral spin symmetry in the putative intermediate phase that large $N_c$ scaling arguments do not resolve.  It has been assumed that in the zero quark masses this phase is chirally restored and the chiral condensate vanishes in the combined large $N_c$ and chiral limits of the theory.  This is an unambiguous notion since chiral symmetry is a symmetry of the theory in that combined limit.  However, chiral spin symmetry is not an explicit symmetry of the theory in that limit, it is a symmetry of only its confining part.  Thus, the issue arises as to whether it is  an exact emergent symmetry that is present in this phase or whether it is an approximate symmetry as it is in the physical world of $N_c=3$ and small but nonzero quark masses.  At this point there is no known dynamical mechanism by which 
 chiral spin symmetry would need to emerge as an exact symmetry in the intermediate phase of the theory in the combined limit.

 At the energy density ${\cal O}(N_c^1)$ the glueballs do not interact, as was discussed above.
 At the same time usual mesons with valence quarks interact strongly and their interaction appears
 at the same order as meson mass in vacuum,  ${\cal O}(N_c^0)$. This would naturally imply  finite
 widths of those mesons, the property, that is observed at $N_c=3$. The intermediate phase could
 thus be considered as a dense system of strongly interacting mesons---with the caveat that these mesons are not the usual ones seen in the vacuum.


\subsection{Quark-gluon plasma phase}

For the purposes of this work, the quark-gluon plasma phase is identified as the phase in which center symmetry is broken and the string tension goes to zero---conditions that become unambiguous at large $N_c$.  The phase is generically characterized by an energy density and pressure that are of order $N_c^2$.  This is precisely the scaling that one would have with a plasma of non-interacting quarks and gluons and would naturally apply to a weakly-interacting plasma as well.  However, the large $N_c$ analysis also implies that the system cannot be a weakly interacting interacting plasma of deconfined quarks and gluons throughout the entire QGP phase.

It is  easy to see that a non-interacting plasma of massless quarks and gluons has the following scaling relations
\begin{equation}
 \epsilon \sim  N_c^2 T^4 \; \; \; P=\frac{\epsilon}{3} \; ; \label{Eq:QGPscal}
\end{equation}
the energy relation holds since the plasma is dominated by  gluons and the number of gluon colors is $N_c^2-1$ each of which contributes an energy density proportional to $T^4$, while the relationship of energy density to pressure is characteristic of non-interacting massless particles.  In a weakly interacting plasma of quarks and gluons one would expect Eq.~(\ref{Eq:QGPscal}) to hold approximately.  Moreover, we know that at high temperatures asymptotic freedom implies that coupling constants will be small so that the the plasma will be weakly interacting and will approach the scaling in Eqs.~\ref{Eq:QGPscal} to  hold.  Note that the running of  the `t Hooft coupling $g N_c^{1/2}$ is described by a $\beta$ function that is independent of $N_c$ for large $N_c$ so that perturbative corrections to the non-interacting plasma become small (in a logarithmic sense) at temperatures of order $\Lambda_{\rm QCD}$ times factors of order $N_c^0$, establishing the scaling in (\Ref{Eq:QGPscal}).

However, regardless of which of the four scenarios enumerated in the introduction is correct, Eq.~(\ref{Eq:QGPscal}) cannot be true in the neighborhood of the phase boundary.  Whether the transition is first-order or second-order, the pressure is continuous across the transition boundary.  Thus, in scenarios \ref{s1} and \ref{s2} the pressure in the QGP phase  just above the transition is of order $N_c^0$ while in the  scenarios \ref{s3} and \ref{s4} the pressure is order $N_c^1$, in either case it is incompatible with Eq.~(\ref{Eq:QGPscal}), even qualitatively.  This is consistent with the generic expectation that the pressure is generically of order $N_c^2$ in the phase; the behavior at the phase boundary is not generic and the coefficient of order $N_c^2$ term simply goes to zero at the boundary.

The important conclusion to draw from this is that the thermodynamics of this phase cannot be interpreted as arising from quarks and gluons moving quasi-freely through this phase. To the extent that the term ``quark-gluon plasma'' is appropriate  throughout this phase, one must conclude that it is a strongly-interacting QGP---at least at the lower end of the phase.  This in turn implies that the mere fact that quarks are ``deconfined'' in the sense that the string tension goes to zero is not sufficient to deduce that quarks and gluons are quasi-free.

\section{The scenarios}

The introduction enumerated four scenarios as to how the qualitative features observed in the world of $N_c=3$ could evolve as $N_c \rightarrow  \infty$.  In this section the plausibility of each of these will be considered.

Scenario \ref{s4}, is interesting in that the intermediate regime seen in lattice studies with $N_c=3$ becomes a distinct phase at large $N_c$.  

A novel feature of that phase is that while the properties of degrees of freedom associated with quarks---as seen through n-point functions of local gauge invariant quark bilinears are radically different from those in the hadron gas phase, all n-point functions of purely gluonic operators are identical to what they were in the hadron gas phase at $T=0$ or in Yang-Mills theory are up to corrections of relative order ${\cal O}(1/N_c)$.  This means that at large $N_c$ ordinary glueballs exist in this phase as they do in the hadron gas phase, and they interact neither with each other or with quark degrees of freedom.

While this behavior is quite natural in the context of large $N_c$ analysis, it is also quite remarkable given the binary way that regimes in QCD are usually viewed.  Traditionally one thinks in terms of a  regime where the system acts as a gas of hadrons or a QGP regime where the quarks and gluons are the effective degrees of freedom.  In this putative intermediate phase glueballs remain as ordinary hadrons but do so in a phase where other hadronic degrees of freedom---the mesons with valence quarks---have their properties radically altered.  This can occur because the $N_c$ scaling of meson with valence quarks and glueballs is characteristically different. 

One might worry that a phase that is so different from the common understanding of how QCD behaves is intrinsically implausible given a half century of experience with the theory.  However, the speculation about the emergence of such a phase is driven by strong numerical evidence that an intermediate regime exists at $N_c=3$ and it is by no means far fetched to speculate that such a regime would become a distinct phase at large $N_c$.  It is also worth noting that something analogous happens at large $N_c$ in the quarkyonic phase of confined but chirally restored matter which is postulated to exist at large baryon chemical potential and has been justified with large $N_c$ arguments\cite{mlp}.

Scenario \ref{s3} is similar to scenario \ref{s4} in having an intermediate phase with the same qualitative  properties noted there.  It differs solely in the order of the transition between hadronic gas matter and the intermediate phase; it is second order in scenario \ref{s3}.  

Assuming that an intermediate phase exists, it is useful to attempt to distinguish between scenarios \ref{s3} and \ref{s4}, {\it i.e.} to determine the order of the  transition between the hadron gas and intermediate phases. It is typically very difficult to determine the order of a transition by theoretical analysis and this problem is no different.  However, there is a plausibility argument suggesting that scenario \ref{s4} is more likely.  

The argument is based on the fact that a second order phase transition has a diverging specific heat at the phase transition temperature. Starting from the hadron gas phase, at large $N_c$ the scaling of the energy density as a function of temperature
is given by Eq.~(\ref{ed}). It is easy to show from Eq. (\ref{ed}), the specific heat cannot diverge unless $N(m)$, the number of hadrons (including spin and flavor degeneracy factors with mass less than $m$) grows exponentially with $m$, {\it i.e.} $N(m) = p(m) \exp
\left ( m/T_H \right )$ where $T_H$ is a parameter known as the Hagedorn temperature\cite{Hag} and $p(m)$ is a subexponential prefactor.  (Note, that while the existence of a Hagedorn spectrum at large $N_c$ is necessary for a second-order transition out of the hadron gas phase, it is not sufficient.  For example, a first-order transition can occur at temperatures below $T_H$.)

Plausibility arguments independent of string-like description of QCD exist that large $N_c$ QCD ought to have a Hagedorn spectrum\cite{Cohen:2009wq,Cohen:2011yx}.  However a natural dynamical explanation for a Hagedorn spectrum is that highly excited states correspond to states with highly excited, and therefore long, flux tubes---which at large $N_c$ cannot break.  To the extent that the flux tubes are much longer than they are wide (as one would expect with highly excited states), they ought to act as strings and string theories are known to have Hagedorn spectra\cite{Zwiebach:2004tj}.  While this string-theory based explanation is a natural picture for why a Hagedorn spectrum could exist and with it the possibility of a second-order transition, string theories should have  Hagedorn spectra for both open strings (mesons) and closed contactible strings (glueballs) with the same $T_H$ for both.  This is  incompatible with scenario  \ref{s3} which would require a Hagedorn spectrum for mesons but also requires that glueballs either do not have a Hagedorn spectrum or have one at a higher temperature.  This suggests that scenario \ref{s4} seems more likely than scenario \ref{s3}

In scenario \ref{s2} the three distinct regimes at $N_c=3$ persist at large $N_c$, but these regimes do not all correspond to distinct phases.  Rather, in this scenario there are two phases: a confined hadronic phase with energy densities and pressures of order $N_c^0$ and a deconfined phase with a vanishing string tension and with energy densities and pressures generically of order $N_c^2$.  In this scenario, hadron-like structures could be  discerned in correlation function at temperatures near the transition even though it is the deconfined nature of the phase; such correlators would reflect an approximate chiral
spin symmetry.  As in the $N_c=3$ world, there could then be  a smooth cross-over to a regime in which these hadron-like structures cease to be discernible as temperature increase.  However, it would
be difficult to explain then why hadron-like structures appear in the deconfined phase at the
same masses as in vacuum, i.e. in the confining phase. This property is clearly seen in lattice
simulations.

Scenario \ref{s1}  is simply that the existence of  an intermediate regime that is seen in lattice studies of the $N_c=3$ world has no analog in the large $N_c$ world.  Such a scenario is uninteresting in the sense that large $N_c$ analysis provides no insight into the phenomena, but this scenario cannot be ruled out on any general grounds. 

In this context it is worth recalling that the degree to which phenomena in the  $N_c=3$ world and the large $N_c$ world correspond, at least qualitatively, varies with the observable being considered. A concrete example of this alluded to earlier: glueballs.  In the large $N_c$ world glueballs must exist and are clearly discernible from mesons while in the physical world of $N_c=3$ there are few if any unambiguous glueball states. Given the possibility that the large $N_c$ world and the physical world can have signinificant qualitative differences, one cannot rule out the possibility that the emergence of an approximate chiral spin symmetry is special to $N_c=3$ and could disappear at large $N_c$. 

The question arises about the width of the intermediate phase. Both deconfinement, $T_d$
and chiral restoration, $T_{ch}$ temperatures are known to scale as $N_c^0$. Consequently
by construction the width of the intermediate phase vanishes as $N_c^{-1}$ in scenarios
1 and 2, while this width is of the order $N_c^0$ in scenarios 3 and 4.

To conclude this discussion it would be important to mention a result
of the large $N_c$ lattice studies of the quark condensate at $T=0$, see \cite{lN1,lN2,lN3,lN4,lN5,DeG1} and references therein.
These studies demonstrate that the quark condensate in the large $N_c$ world is quite
close to its value at $N_c=3$. This suggests that  for the
 spontaneous breaking of chiral symmetry the $1/N_c$ corrections are small and the physics of this phenomenon in the large
 $N_c$ world and in the real world $N_c=3$ is likely to be similar.  Thus, it is plausible that
  the chiral symmetry restoration phase transition in
 the large $N_c$ world---the temperature of the phase transition (in the chiral limit)---
 should be close to the temperature of the chiral restoration phase
 transition at $N_c=3$. The latter temperature is known from the lattice calculations
 to be around $T \sim 130$ MeV \cite{Karsch}.

 In the large $N_c$ world the quark loops are suppressed and the physics of
 confinement should be similar to that of  pure Yang-Mills theory
 or in quenched QCD. In pure Yang-Mills and in quenched QCD the deconfinement phase
 transition of the first order is known to happen at $T \sim 270 - 300$ MeV for the case of $N_c=3$.  
Thus, it is plausible that in the large $N_c$ world there exist  two distinct temperatures, one for the
chiral restoration phase transition (around $T^{large N_c}_{ch} \sim 130 $ MeV)
 and for the  deconfinement transition at a temperature around 300 MeV. 
 This qualitative prediction
 could and should be checked in large $N_c$ lattice simulations. If correct, it would be
 direct verification that either scenario 3 or scenario 4 is realised.

\section{Conclusions}

 In this paper we have considered the QCD phase diagram at zero baryon
chemical potential and established $N_c$ scalings of different regimes.
This analysis is  potentially important since in the large $N_c$ limit the center symmetry and confinement notions are well defined and consequently
allow one to unambiguously discuss phases that are characterized by confinement or
deconfinement as well as with explicit chiral symmetry or its spontaneous breaking.  We have identified four scenarios as to how the qualitative features seen in the physical world can be generalized to large $N_c$.  Lattice calculations at large $N_c$ would help to distinguish which of these is correct.  
While lattice studies for full QCD with $N_c >3$ have been done\cite{DeG2} the values of $N_c$ have been modest ($N_c \le 5$) and calculations at larger $N_c$ would be needed in full QCD to determine which scenario is correct, because the quark loops
get suppressed slowly, as $1/N_c$. Since the chiral limit and the large $N_c$ limit
do not commute, for the present issue the large $N_c$ limit should be taken first.  \footnote{It is rather straightforward to take the large $N_c$ limit for
the purposes of the present paper. In the large $N_c$ limit
the quark loops get suppressed and the full QCD becomes
quenched. Consequently to find the deconfinement temperature
of full QCD in the large Nc limit one needs to find the deconfinement temperature
in pure glue theory at three different $N_c=3,5,7$ and extrapolate to $N_c=\infty$
\cite{LTW,lN2}.
The same is true with the chiral restoration temperature. Given the
pure glue configurations at $N_c=3,5,7$ one solves the Dirac eigenvalue problem
and obtains the quark condensate from the density of the near zero modes
via the Banks-Casher relation. Increasing the temperature one extracts the chiral
restoration temperature, at which the quark condensate vanishes, for a given $N_c$.
Then one extrapolates to $N_c=\infty$.
If at $N_c=\infty$ one finds $T_{ch} < T_d$, the existence of the intermediate phase
will be proven.}

Given what is known about critical temperatures in the real world and on the magnitude of the quark condensate at $T=0$ at large $N_c$,  we have
argued that scenarios in which there are distinct  temperatures of chiral and deconfinement phase transitions  in the large $N_c$ world are plausible.

We note that the analysis here could be extended to the inclusion of chemical potentials.  Probably the simplest extension is to the inclusion of an isospin chemical (since unlike baryon chemical potentials which couple to degrees of freedom with a minimum energy of order $N_c$ the isospin chemical potential couples to excitations with a minimum excitation of $N_c^0$).  The isospin chemical potential also has the virtue that it can be simulated in lattice studies without a sign problem\cite{Brandt:2017oyy,Kogut:2002zg}.  The study of large $N_c$ QCD with a nonzero chemical potential can lead to the exploration of a very rich phase diagram.\footnote{Since this paper has been completed, some
progress has been achieved with respect to what scenario is correct. In partucular, an analysis of the $N_c$ scaling of the fluctuations of conserved charges and of Polyakov loop together with the lattice data for $N_c=3$ suggests
that either scenario 3 or 4 should be correct \cite{fluc}.  A solvable large $N_c$ chirally symmetric and confining model in 3+1 dimensions, that is similar
to 't Hooft model (large $N_c$ QCD in 1+1 dimensions), demonstrates a manifetly
confining and chirally symmetric phase with approximate chiral spin symmetry
above the chiral restoration temperature \cite{GNW}.}

\section*{Acknowledgements }
The authors thank Owe Philipsen for interesting and insightful comments. They also thank Aleksey Cherman for noting the issue of the Coleman-Witten theorem.  The work of TDC was supported in part by the the U.S. Department
of Energy, Office of Nuclear Physics under Award Number
DE-FG02-93ER40762.

\end{document}